\documentclass[conference]{IEEEtran}
\usepackage{graphicx}
\usepackage{amssymb}
\usepackage{amsfonts}
\usepackage{amsmath}
\IEEEoverridecommandlockouts
\newtheorem{Theorem}{Theorem}
\newtheorem{Proposition}{Proposition}
\newtheorem{Definition}{Definition}
\newtheorem{Example}{Example}
\ifCLASSINFOpdf
\else
\fi
\begin{document}
\title{On the Performance of 1-level LDPC Lattices}
\author{\IEEEauthorblockN{Mohammad-Reza~Sadeghi}
\IEEEauthorblockA{Faculty of Mathematics and Computer Science,\\
Amirkabir University of Technology\\
Email: msadeghi@aut.ac.ir}
\and
\IEEEauthorblockN{Amin~Sakzad}
\IEEEauthorblockA{Department of ECSE,\\
Monash University, Victoria, Australia\\
E-mail: amin.sakzad@monash.edu}
}

\maketitle
\begin{abstract}
The low-density parity-check (LDPC) lattices perform very well in high dimensions under generalized min-sum iterative decoding algorithm. In this work we focus on $1$-level LDPC lattices. We show that these lattices are the same as lattices constructed based on Construction A and low-density lattice-code (LDLC) lattices. In spite of having slightly lower coding gain, $1$-level regular LDPC lattices have remarkable performances. The lower complexity nature of the decoding algorithm for these type of lattices allows us to run it for higher dimensions easily. Our simulation results show that a $1$-level LDPC lattice of size $10000$ can work as close as $1.1$ dB at normalized error probability (NEP) of $10^{-5}$.This can also be reported as $0.6$ dB at symbol error rate (SER) of $10^{-5}$ with sum-product algorithm.
\end{abstract}
\section{Introduction}
\PARstart{P}{oltyrev}~\cite{polytrev} suggests and investigates coding without restriction for infinite arrays such as lattices on the AWGN channel. That is a communication without power constraints. In this communication scheme, instead of the coding rate and capacity two new concepts of normalized logarithmic density (NLD) and generalized capacity $C_{\infty}$ are defined. Forney {\em et al.} ~\cite{forneyspherebound} showed the existence of sphere-bound-achieving and capacity-achieving lattices via Construction D lattices theoretically. He also established the concept of volume-to-noise (VNR) ratio as a parameter for measuring the efficiency of lattices. Therefore, generalized capacity for lattices means: there exists a lattice of high enough dimension $n$ that enables transmission with arbitrary small error probability whenever VNR approaches $1$. In addition, it has been shown in~\cite{polytrev} that, this error probability is bounded away from zero when $\mbox{VNR}<1$. A capacity-achieving lattice can raise to a capacity-achieving lattice code by selecting a proper shaping region~\cite{erez,urbanke}.

The search for sphere-bound-achieving and capacity-achieving lattices and lattice codes has begun with~\cite{sadeghi}. LDPC lattices are those that have sparse parity check matrices. These lattices were introduced first by Sadeghi {\em et al.}~\cite{sadeghi}. In this class of lattices, a set of nested LDPC codes are used to generate lattices with sparse parity check matrices. Another class of LDPC lattices, so-called LDLC lattices introduced and investigated in~\cite{loeligerLDLC,LDLC}. Turbo lattices employed Construction D along with Turbo codes to achieve capacity gains~\cite{sakzad10, sakzadmanuscript}.

An extended version of {\em Progressive Edge Graph} (PEG) algorithm, called {\em extended PEG} (E-PEG) algorithm~\cite{sadeghi} can be employed to construct the Tanner graph of regular LDPC lattices. Since presence of short cycles can hurt the performance of lattices~\cite{sakzad}, such construction provides us the Tanner graphs for lattices with high girth. This will have an impact on the excellent performance of LDPC lattices.

In the present work, we introduce and investigate 1-level regular LDPC lattices~\cite{sadeghi}. We show that these lattices are equivalent to lattices constructed based on Construction A. We also provide a relationship between these lattices and the well-known LDLC lattices. Finally, we provide experimental results to reveal the effectiveness of 1-level regular LDPC lattices in terms of fundamental coding gain and error probability.

\section{Preliminaries on Lattices}~\label{Preliminaries}
In order to make this work self-contained, general notations and basic definitions are given next.
\subsection{Lattices}~\label{A.Lattices}
A discrete, additive subgroup $\Lambda$ of the $m$-dimensional real space $\mathbb{R}^m$ is called a \emph{lattice}. We denote the length of the shortest nonzero vector of $\Lambda$ by $d_{\min}(\Lambda)$. Indeed, $d_{\min}(\Lambda)$ refers to the \emph{minimum distance} between different points of the lattice. Every lattice $\Lambda$ has a basis $\mathcal{B}=\{{\bf b}_1,\ldots,{\bf b}_n\}\subseteq\mathbb{R}^m$ where every ${\bf x}\in\Lambda$ can be represented as an integer linear combination of vectors in $\mathcal{B}$. Let
$\mbox{span}(\Lambda)$ be
$$\{\alpha_1{\bf b}_1+\cdots+\alpha_n{\bf b}_n\colon\alpha_i\in\mathbb{R}\},$$ then a {\em Voronoi cell} $\nu({\bf x})$ can be defined as the set of those points of $\mbox{span}(\Lambda)$ that are at least as close to the point ${\bf x}$ as to any other lattice point. The set of all vectors in $\mbox{span}(\Lambda)$ where their inner product with all the vectors of $\Lambda$ is in $\mathbb{Z}$ forms another lattice, denoted by $\Lambda^\ast$, called the {\em dual lattice} of $\Lambda$.
Define the {\em coding gain} of $\Lambda$ as:
\begin{equation}~\label{codinggain}
\gamma(\Lambda)=\frac{d^2_{\min}(\Lambda)}{(\det(\Lambda))^{\frac{2}{n}}},
\end{equation}
where the $\det(\Lambda)$ is the {\em volume} of the lattice. The parameter $\det(\Lambda)^{\frac{2}{n}}$ refers to the~\emph{normalized volume} of an $n$-dimensional lattice $\Lambda$ ~\cite{forneyspherebound}. This volume may be regarded as the volume of $\Lambda$ per two dimensions. Suppose that the points of a lattice $\Lambda$ are sent over an AWGN channel without restrictions with noise variance $\sigma^2$. The parameter
\begin{equation}~\label{VNR}
{\mbox {VNR}}=\frac{\det(\Lambda)^{\frac{2}{n}}}{2\pi e\sigma^2},
\end{equation}
is the \emph{volume-to-noise ratio} (VNR) of lattice $\Lambda$. When $n$ is large enough, the parameter VNR can be interpreted as the ratio of the normalized volume of the lattice to the normalized volume of a noise sphere of squared radius $n\sigma^2$. This is also defined as generalized signal-to-noise ratio (SNR) in~\cite{sadeghi} and $\alpha^2$ in~\cite{forneyspherebound}.

Let the vector $\textbf{x}$ in $\Lambda$ be transmitted on the unconstrained AWGN channel, then the received vector $\textbf{r}$ can be written as $\textbf{r}=\textbf{x}+\textbf{e}$
where $\textbf{e}=(e_1,\ldots,e_n)$ is in the Euclidean space and its components are independently and identically distributed (i.i.d.) Gaussian random variables with zero mean and variance $\sigma^2$. Thus the probability of correct decoding is given by
\begin{equation}~\label{pe}
P_c=\frac{1}{(\sigma\sqrt{2\pi})^n}\int_{\nu(\textbf{x})}e^{\frac{-\|{\bf t}\|^2}{2\sigma^2}}d{\bf t},
\end{equation}
where $\|{\bf x}\|$ is the Euclidean norm of ${\bf x}$. Due to the geometric uniformity of lattices and without loss of generality we suppose that the vector ${\bf 0}$ is sent and the vector $\textbf{r}$ is received. Then the components of $\textbf{r}$ are i.i.d. Gaussian distributed random variables with zero mean and variance $\sigma^2$. The probability of error under a maximum likelihood decoder for $\Lambda$ is the same as the probability that a white Gaussian $n$-tuple ${\bf r}$ with noise variance $\sigma^2$ fall outside the Voronoi region $\nu({\bf 0})=\nu$, {\em i.e.} $P_e=1-P_c$. The {\em normalized error probability (NEP)}, $P_e^\ast$, of an $n$-dimensional lattice $\Lambda$ is the error probability per two dimensions, {\em i.e.} $P_e^\ast=\frac{2}{n}P_e$~\cite{forneyspherebound,tarokh}.
\subsection{Geometric structure of lattices}~\label{GeometricStructureofLattices}
We suppose that the $n$-dimensional lattice $\Lambda'$ is a sublattice of an $n$-dimensional lattice $\Lambda$, and the lattice $\Lambda'$ has a basis along the orthogonal subspaces
$\mathcal{S}=\{\mathcal{W}_i\}^n_{i=1}$ with $\dim(\mathcal{W}_i)=1$. For every point ${\bf x}$ in $\Lambda$, the label sequence is $g({\bf x})=(g_1({\bf x}),\ldots,g_n({\bf x}))$
where $g_i({\bf x})=\mathrm{P}_{\mathcal{W}_i}({\bf x})+\Lambda_{\mathcal{W}_i}$ and $\mathrm{P}_{\mathcal{W}_i}(\Lambda)$ is the projection onto the vector space $\mathcal{W}_i$ and the cross section $\Lambda_{\mathcal{W}_i}$ is $\Lambda\cap \mathcal{W}_i$. The {\em label code} of $\Lambda$ is then the set of all possible label sequences $\mathcal{L}=g(\Lambda)=\{g({\bf x}):{\bf x} \in \Lambda\}$. It is a group code over the alphabet $\mathcal{G}=\mathbb{Z}_{g_1}\times\cdots\times\mathbb{Z}_{g_n}$, i.e. $\mathcal{L}\subseteq \mathcal{G}$. For every ${\bf a}, {\bf c}\in \mathcal{G}$, one can define~\cite{banihashemi} the inner product $({\bf a},{\bf c})$ by:
\begin{equation}~\label{innerproduct}
({\bf a},{\bf c}):=\frac{a_1c_1}{g_1}+\cdots+\frac{a_nc_n}{g_n}\pmod{\mathbb{Z}}
\end{equation}
where the operations like multiplications and divisions are done in $\mathbb{R}$. For every subgroup $\mathcal{L}\subseteq\mathcal{G}$, the dual of $\mathcal{L}$, is defined by
$\mathcal{L}^\ast=\{{\bf c}\in \mathcal{G}\mid({\bf c},{\bf a})=0 ,~\forall {\bf a}\in \mathcal{L}\}$. Based on the above discussion, over the coordinate system  $\mathcal{S}=\{\mathcal{W}_i\}_{i=1}^n$,
a lattice $\Lambda$ can be decomposed as~\cite{banihashemi,sakzad}
\begin{equation}~\label{beutifulformula}
\Lambda=\mathbb{Z}^n{\bf C}(\Lambda)+\mathcal{L}{\bf P}(\Lambda)
\end{equation}
where $\mathcal{L}$ is a group code over $\mathbb{Z}_{g_1}\times\mathbb{Z}_{g_2}\times\cdots\times\mathbb{Z}_{g_n}$ and
$${\bf C}(\Lambda)=\mbox{diag}(\det(\Lambda_{\mathcal{W}_1}),\dots,\det(\Lambda_{\mathcal{W}_n}))$$
$${\bf P}(\Lambda)=\mbox{diag}(\det (\mathrm{P}_{\mathcal{W}_1}(\Lambda)),\dots,\det (\mathrm{P}_{\mathcal{W}_n}(\Lambda)))$$
where $\mbox{diag}(\cdots)$ denotes a diagonal matrix.
The decomposition~(\ref{beutifulformula}) means that a vector ${\bf v}\in \mathbb{R}^n$ is in $\Lambda$ if and only if it can be written as ${\bf v}={\bf z}{\bf C}(\Lambda)+{\bf c}{\bf P}(\Lambda)$ for some ${\bf z}\in \mathbb{Z}^n$ and ${\bf c}\in \mathcal{L}$. It is also shown~\cite{banihashemi} that $\Lambda^\ast$ can be represented in terms of $\mathcal{L}^\ast$, ${\bf P}(\Lambda)$ and ${\bf C}(\Lambda)$ as follows:
\begin{equation}~\label{eq:duallattice}
\Lambda^\ast=\mathbb{Z}^n{\bf P}^{-1}(\Lambda)+\mathcal{L}^\ast {\bf C}^{-1}(\Lambda).
\end{equation}
In other words, let $\mathcal{L}$ be the label code of $\Lambda$, then the dual of $\mathcal{L}$, denoted by $\mathcal{L}^\ast$, is the label code of $\Lambda^\ast$. It has to be noted that $g_i^\ast$, the $i$-th alphabet size of $\mathcal{L}^\ast$, equals $g_i$, the $i$th alphabet size of $\mathcal{L}$, for $1\leq i\leq n$.

The above representations for $\Lambda$ and $\Lambda^\ast$ have a practical consequence that we can derive a generating set of lattice $\Lambda^\ast$ based on a generating set for $\mathcal{L}^\ast$ and vice versa. More specifically, assume that $\{{\bf v}^\ast_1,\ldots,{\bf v}^\ast_n\}$ be a generating set for $\Lambda^\ast$, then a generating set $\{{\bf c}^\ast_1,\ldots,{\bf c}^\ast_r\}$ can be derived easily. To obtain the $j$--th coordinate of ${\bf c}^\ast_i$, it is sufficient to divide the $j$th coordinate of ${\bf v}^\ast_i$ by $(j,j)$--th entry of ${\bf C}^{-1}(\Lambda)$ which is $1/\det(\Lambda_{W_j})$ and then evaluate the result module $g_j$. On the other hand, if $\{{\bf c}^\ast_1,\ldots,{\bf c}^\ast_r\}$ be a generating set for $\mathcal{L}^\ast$, then $\mathcal{L}^\ast {\bf C}^{-1}(\Lambda)$ along with the following vectors form a generating set of $\Lambda^\ast$:

{\small
$$
\begin{array}{lrrrrrrr}
(\hspace{-1cm}&\frac{1}{\det (P_{W_1}(\Lambda))},&0,&0,&\ldots,&0,&0)\\
(\hspace{-1cm}&0,&\frac{1}{\det (P_{W_2}(\Lambda))},&0,&\ldots,&0,&0)\\
&&&\ldots&&\\
(\hspace{-1cm}&0,&0,&0,&\ldots,&0,&\frac{1}{\det (P_{W_n}(\Lambda))}).
\end{array}
$$}
To construct a Tanner graph for group code $\mathcal{L}$~\cite{banihashemi}, we use the following set of check equation constraints, which fully describes the codewords ${\bf c}=(c_1,\ldots,c_n)$:
\begin{equation}~\label{secondinnerproduct}
\sum_{i=1}^n \frac{c_{ki}^\ast
c_i}{g_i}\in\mathbb{Z},~~~~~~~~~1\leq k\leq r
\end{equation}
where $\mathcal{C}^\ast=\{{\bf c}_1^\ast,\ldots,{\bf c}_r^\ast\}$ is a generating set for $\mathcal{L}^\ast$. A symbol node represents a symbol while a check node represents a check equation in \eqref{secondinnerproduct}. The check nodes and symbol nodes are connected by edges. In fact, we put an edge between $j$--th symbol node and $i$--th check node if and only if ${\bf c}^\ast_{ij}\neq0$. A Tanner graph for a lattice $\Lambda$ is then the Tanner graph of its label code (based on construction on $\mathcal{L}^\ast$).
\section{LDPC lattices}~\label{LDPC Lattices}
There exist many ways to construct lattices based on codes~\cite{conway}. Here we mention two of them. The first one is Construction A and the other one is Construction D'.
\subsection{Construction A versus Construction D'}~\label{ConstructionAversusConstructionD'}
Assume that $\mathcal{C}$ is a linear code over $\mathbb{Z}_p$ where $p$ is a prime number, i.e. $\mathcal{C}\subseteq\mathbb{Z}_{p}^n$. Let $d_{\min}$ denotes the minimum distance of $\mathcal{C}$. A lattice $\Lambda$ constructed based on Construction A~\cite{conway} can be derived from $\mathcal{C}$ by:
\begin{equation}\label{constA}
\Lambda=p\mathbb{Z}^n+\epsilon\left(\mathcal{C}\right),
\end{equation}
where $\epsilon\colon\mathbb{Z}_p^n\rightarrow\mathbb{R}^n$ is the embedding function which sends a vector in $\mathbb{Z}_p^n$ to its real version. In this work, we are particularly interested in binary codes and lattices with $p=2$.

Construction D' for lattices~\cite{conway} is a good tool for lattice construction based on LDPC codes. Let $\mathcal{C}_0\supseteq \mathcal{C}_1\supseteq \cdots \supseteq \mathcal{C}_a$ be a set of nested linear block codes, where $\mathcal{C}_\ell\left[n,k_\ell,d_{\min}^\ell \right]$, for $0\leq \ell\leq a$. Let $\{{\bf h}_1,\ldots ,{\bf h}_n\}$ be a basis for $\mathbb{F}_2^n$, where the code $\mathcal{C}_\ell$ is formed by the $r_\ell=n-k_\ell$ parity check vectors ${\bf h_1},\ldots ,{\bf h}_{r_\ell}$.
If we consider vectors ${\bf h}_j$, for $1\leq j \leq n$, as real vectors with elements $0$ or $1$ in $\mathbb{R}^n$. The new lattice $\Lambda$ includes all vectors ${\bf x}\in \mathbb{Z}^n$ satisfying the following modular equations ${\bf h}_j\cdot {\bf x}\equiv 0 \pmod{2^{\ell+1}}$, for $0\leq \ell\leq a$ and $r_{a-\ell-1}+1\leq j\leq r_{a-\ell}$. The number $a+1$ is called the level of the construction.

By multiplying the above congruences by proper powers of $2$, one restates Construction D'~\cite{sadeghi}. Indeed, ${\bf x}$ is in $\Lambda$ if and only if  ${\bf H}{\bf x}^T=0 \pmod{2^{a+1}}$ where
\begin{equation}~\label{H}
{\bf H}=[{\bf h}_1,\ldots,{\bf h}_{r_0},\ldots,2^a{\bf h}_{r_{a-1}+1},\ldots,2^a{\bf h}_{r_{a}}]^T.
\end{equation}
Then ${\bf H}$ is called the parity check matrix of the lattice $\Lambda$. The lattice $\Lambda$ built by Construction D' and associated to this ${\bf H}$ is called an {\em $(a+1)$-level lattice}. Hence the Tanner graph of these lattices can be constructed based on their parity check matrices ${\bf H}$. It can be shown~\cite{sadeghi} that the volume of an $(a+1)$-level lattice $\Lambda$ is
\begin{equation}~\label{det}
\det(\Lambda)=2^{\sum_{\ell=0}^{a}r_\ell}.
\end{equation}
Also the minimum distance of $\Lambda$ satisfies the following bounds
\begin{equation}~\label{minimum distance}
\min_{0\leq \ell\leq a}\left\{4^\ell d_{\min}^{a-\ell}\right\}\leq d_{\min}^2(\Lambda)\leq4^{a+1}.
\end{equation}
A lattice is {\em regular} if its Tanner graph, or its corresponding parity check matrix ${\bf H}$, be $(d_s,d_c)$-regular. In other words, the number of nonzero elements in all rows and columns are the same and are equal to $d_s$ and $d_c$, respectively. Now the following example reveals the nature of a lattice constructed based on Construction D'. \begin{Example}
Let $a=2$ and $\mathcal{C}_0$, $\mathcal{C}_1$ and $\mathcal{C}_2$ are three nested codes. The sets $\{0110\}$, $\{0110,~1010\}$ and $\{0110,~1010,~1100\}$, are the basis for their dual codes respectively. Therefore, we have ${\bf h}_1=(0,1,1,0)$, ${\bf h}_2=(1,0,1,0)$ and ${\bf h}_3=(1,1,0,0)$. Let
$${\bf H}=\left[
\begin{array}{cccc}
0&1&1&0\\
2&0&2&0\\
4&4&0&0
\end{array}\right].$$
Then ${\bf x}\in\mathbb{Z}^n$ is in $\Lambda$ if and only if ${\bf H}{\bf x}^T\equiv{\bf 0}\pmod{2^3}$. This is a $3$-level, $(2,2)$-regular lattice.
\end{Example}
A lattice $\Lambda$ formed by Construction D' is called {\em regular LDPC lattice} if its parity check matrix ${\bf H}$ is a sparse $(d_s,d_c)$-regular matrix~\cite{sadeghi}. It is easy to see that if the nested codes $\mathcal{C}_\ell$ are LDPC codes then the obtained lattice is a LDPC lattice.

The Extended Progressive Edge Growth Algorithm (E-PEG) was introduced and employed to build regular bipartite graph~\cite{sadeghi} (regular LDPC lattices). For any $0\leq \ell\leq a$, $d_{s}^\ell$ denotes the degree of the symbol node $s$ in the parity check matrix of $\mathcal{C}_\ell$. Also for every check node $c$, the notation $d_{c}$ refers to the degree of that node. Suppose that for $0\leq \ell\leq a$, $d_s^\ell=\ell+2$ for every symbol node $s$ and select a divisor $d_c=2^{a+1}$ of $n$ such that $d_c>d_s$. Also let $r_\ell$ be such that
\begin{equation}~\label{eq:rlregularlattice}
r_\ell=\frac{\ell+2}{2^{a+1}}n
\end{equation}
An $(a+1)$-level lattice with these parameters is called an {\em $(a+2,2^{a+1};a+1)$ regular LDPC lattice}~\cite{sadeghi}. These classes of lattices are denoted by $\Lambda^{a+1}_n$. For sufficiently large $n$, $\Lambda^{a+1}_n$ is an LDPC lattice with coding gain~\cite{sadeghi} $$\gamma(\Lambda^{a+1}_n)=4^{a+1-\frac{(a+1)(a+2)}{2^{a+2}}}.$$
Now, the E-PEG algorithm is employed to construct an {\em $(a+2,2^{a+1};a+1)$ regular LDPC lattice}. Indeed, the E-PEG algorithm gets $a,n$ and $r_\ell$, $0\leq \ell\leq a$,
as its inputs and returns an $(a+1)$-level Tanner graph. This Tanner graph corresponds to an $(a+2,2^{a+1};a+1)$ regular LDPC lattice. As we see, if the initial inputs of the E-PEG are chosen appropriately, then a set of regular Tanner graphs and consequently a class of regular LDPC lattices would be constructed. An example for a $(3,4;2)$ regular lattice is given next.
\begin{Example}
 Consider the $(3,4;2)$ regular lattice
$$\left[
\begin{array}{cccccccccccccccc}
  1&0&0&0&1&0&0&0&1&0&0&1&0&0&0&0\\
  1&0&0&0&0&1&0&0&0&1&0&0&1&0&0&0\\
  0&1&0&0&1&0&0&0&0&1&0&0&0&1&0&0\\
  0&1&0&0&0&0&1&0&1&0&0&0&0&0&1&0\\
  0&0&1&0&0&0&1&0&0&0&1&0&0&1&0&0\\
  0&0&1&0&0&0&0&1&0&0&0&1&0&0&1&0\\
  0&0&0&1&0&1&0&0&0&0&1&0&0&0&0&1\\
  0&0&0&1&0&0&0&1&0&0&0&0&1&0&0&1\\
  2&0&0&0&0&0&2&2&0&0&0&0&0&2&0&0\\
  0&2&0&0&0&2&0&0&0&0&2&2&0&0&0&0\\
  0&0&2&0&2&0&0&0&0&0&0&0&2&0&2&0\\
  0&0&0&2&0&0&0&0&2&2&0&0&0&0&0&2\\
\end{array}
\right]
$$
Here $a=1$, $d_{c_i}=4$ for $1\leq i\leq12$, $d_s^0=2$ and $d_s^1=3$.
\end{Example}
Now we disclose the relationship between a $1$-level lattice constructed based on Construction D' and a Construction A lattice.
\begin{Proposition}~\label{th:equivalenD'A}
Let $\Lambda_0$ be generated with Construction D' with linear code $\mathcal{C}_0$ and $a=0$, then it is equal to a lattice $\Lambda_1$ constructed following Construction A using the same underlying code $\mathcal{C}_0$.
\end{Proposition}
\begin{IEEEproof}
Suppose that $\Lambda_0$ be the lattice generated by $\mathcal{C}_0$ based on Construction D' and $a=0$. Also let us assume that $\Lambda_1=2\mathbb{Z}^n+\epsilon\left(\mathcal{C}_0\right)$ is the lattice constructed following Construction A. The vector ${\bf x}\in \mathbb{Z}^n$ is in $\Lambda_0$ if it satisfies the congruences ${\bf h}_j\cdot {\bf x}\equiv 0 \pmod{2}$ for $1\leq j\leq r_{0}$. Every ${\bf x}\in \Lambda_1$ can be represented as ${\bf x}={\bf y}+{\bf c}$ for ${\bf y}\in2\mathbb{Z}^n$ and ${\bf c}\in \mathcal{C}_0$. Therefore for $1\leq j\leq r_{0}$ we get $${\bf h}_j\cdot {\bf x}\equiv {\bf h}_j\cdot ({\bf y}+{\bf c})\equiv0\pmod{2},$$
since ${\bf y}\in2\mathbb{Z}^n$ and ${\bf h}_j\cdot {\bf c}\equiv0\pmod{2}$. It concludes that $\Lambda_1\subseteq\Lambda_0$. If ${\bf x}\in \Lambda_0$, then ${\bf h}_j\cdot {\bf x}\equiv 0 \pmod{2}$ for $1\leq j\leq r_{0}$. By subtracting an appropriate ${\bf z}\in2\mathbb{Z}^n$ from ${\bf x}$, we get ${\bf x}-{\bf z}\in\{0,1\}^n$ or $\epsilon^{-1}\left({\bf x}-{\bf z}\right)\in\mathbb{Z}_2^n$. Since ${\bf h}_j\cdot ({\bf x}-{\bf z})\equiv 0 \pmod{2}$ for $1\leq j\leq r_{0}$, the vector $\epsilon^{-1}\left({\bf x}-{\bf z}\right)$ should be in $\mathcal{C}_0$. Thus $\epsilon^{-1}\left({\bf x}-{\bf z}\right)={\bf c}_0$ for some ${\bf c}_0\in\mathcal{C}_0$ and hence $\Lambda_0\subseteq\Lambda_1$. It turns out that $\Lambda_0=\Lambda_1$.
\end{IEEEproof}
\begin{Definition}
A 1-level LDPC lattice $\Lambda$ is a lattice constructed based on Construction D' along with binary linear LDPC code $\mathcal{C}$ as its underlying code. Equivalently, ${\bf x}\in\mathbb{Z}^n$ is in $\Lambda$ if ${\bf H}{\bf x}^T=0 \pmod{2}$ where ${\bf H}=[{\bf h}_1,\ldots,{\bf h}_{r_0}]^T$ is the parity check matrix of $\mathcal{C}$.
\end{Definition}
Based on Proposition~\ref{th:equivalenD'A}, it is clear that a 1-level LDPC lattice $\Lambda$ can be constructed by Construction A with underlying code $\mathcal{C}$ and it also can be represented as $\Lambda=2\mathbb{Z}^n+\epsilon\left(\mathcal{C}\right)$.
\subsection{LDPC lattices are the same as LDLC}~\label{LDPCLatticesareLDLC}
Low-Density Lattice Codes (LDLC) were introduced in~\cite{LDLC}. These lattices have attracted attentions recently~\cite{loeligerLDLC}. An $n$ dimensional LDLC is an $n$-dimensional
lattice with a nonsingular generator matrix ${\bf G}$ for which the parity check matrix ${\bf H}=\left({\bf G}^{-1}\right)^T$ is sparse. In this subsection, we show that LDPC lattices are LDLC.
\begin{Theorem}\label{th:LDLCLDPC}
Let $\Lambda=\Lambda_n^{a+1}$ be a regular LDPC lattice built by Construction D'. Then $\Lambda$ is an LDLC.
\end{Theorem}
\begin{IEEEproof}
It is shown in~\cite{sadeghi} that if we choose $\mathcal{C}_0$ such that $d_{\min}^0\geq2$, then $\Lambda_{\mathcal{W}_i}=2^{a+1}\mathbb{Z}$ and $\mathrm{P}_{\mathcal{W}_i}(\Lambda)=\mathbb{Z}$, for every $1\leq i\leq n$. Hence $\Lambda^\ast=\mathbb{Z}^n+\frac{1}{2^{a+1}}\mathcal{L}^\ast$. It means that $g_i=2^{a+1}$ and in the view of representation~(\ref{eq:duallattice}), we have ${\bf P}^{-1}(\Lambda)={\bf I}_n$ and ${\bf C}^{-1}(\Lambda)=\frac{1}{2^{a+1}}{\bf I}_n$. In the case for which $g_i=2^{a+1}$ the row vectors of ${\bf H}$ as in~(\ref{H}) constitute the generator for $\mathcal{L}^\ast$. In other words, ${\bf H}$ is the generator matrix of $\mathcal{L}^\ast$, the dual of label code $\mathcal{L}$ of $\Lambda$. Since we have constructed LDPC lattices such that ${\bf H}$ is sparse. So a generator matrix for $\Lambda^\ast$ can be obtained using the method described in~\ref{GeometricStructureofLattices}. Thus the row vectors of ${\bf H}$ along with row vectors of ${\bf I}_n$ form a generating set for $\Lambda^\ast$. This implies that regular LDPC lattices $\Lambda_n^{a+1}$ are LDLC.
\end{IEEEproof}
\section{Numerical Analysis of 1-level LDPC lattices}~\label{Simulations}
\subsection{Generalized min-sum algorithm}~\label{generalizedmin-sumalgorithm}
This algorithm is the generalization of min-sum algorithm. It can be employed to solve CVP. An explanation of this algorithm is given in~\cite{sadeghi}. In order to solve CVP, given an input vector ${\bf y}$, this algorithm finds the closest  lattice vector ${\bf x}$ to ${\bf y}$ by finding the label codeword of the coset which ${\bf x}$ belongs to. The generalized min-sum is a message passing iterative algorithm. In each iteration two steps are performed: (1) {\em symbol node operation} and (2) {\em check node operation}. Also in each iteration algorithm does a {\em hard decision}. For further details, we refer the reader to~\cite{sadeghi}.

\subsection{Simulation results and comments for $1$-level LDPC lattices}~\label{LDPCLatticesareLDLC}
It is known that maximum likelihood decoder on the AWGN for codes is minimum distance decoder and maximum likelihood decoder on the AWGN without restriction for lattices is Euclidean minimum distance decoder. Now assume that we are going to use the above generalized min-sum algorithm in order to decode a 1-level LDPC lattice. In this case for $1\leq i\leq n$, we have $g_i=2$, $\det(\Lambda_{\mathcal{W}_i})=2$ and $\det(\mathrm{P}_{\mathcal{W}_i}(\Lambda))=1$. It has two consequences. First, the decoding complexity, which is provided in~\cite{sadeghi}, is much lower than the case for 2-level ones. Second, with the above parameters the initial weights of the generalized min-sum algorithm are equal to the squared distance between the $i$-th symbol of the received word and different cosets of the $i$-th coordinate, which are only two cosets. In fact, these weights play the same role as the log-likelihood costs do in the original min-sum algorithm. It means that in the case of 1-level LDPC lattice our generalized min-sum algorithm turns to be the original min-sum algorithm which uses Euclidean distance instead of the negative log-likelihood costs. On the other hand, the squared distances between the elements of the label group and each coordinate of the received word are bounded above by $1$, instead of $4$ in the $2$-level case. Furthermore, preliminary results show that by using $1$-level LDPC lattices, our decoder stops after $10$ to $15$ iterations. Since we are adding the costs during the decoding process, this causes generation of large numbers in the case of $2$-level lattices in contrast with $1$-level lattices. These two advantages allow the algorithm runs for much higher dimensions.

The simulation results of $1$-level regular LDPC lattices are given in Fig. 1.
\begin{figure}[htb]
\begin{center}
\includegraphics[width=7.5cm]{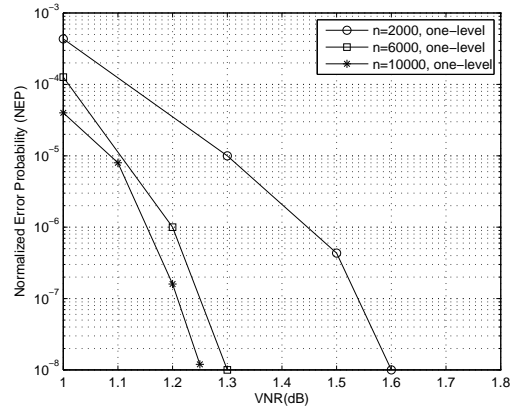}
\caption{Performance of $2000$, $6000$ and $10000$-dimensional $1$-level regular LDPC lattices.}
\end{center}
\end{figure}
The graphs show that a gain of $1$ dB can be obtained if we employ $1$-level LDPC lattices rather than $2$-level LDPC lattices~\cite{sadeghi}. Also $1$-level regular LDPC lattices of sizes $n=6000$ and $n=10000$, at NEP of $10^{-5}$, achieves $\alpha^2=1.15$ dB and $\alpha^2=1.1$ dB away from capacity respectively. At the moment, the performance of these lattices are being evaluated under sum-product algorithm as well. The primary results show that at $n=10000$ these lattices may work as close as $0.67$ dB at SER $10^{-5}$. This is a normal achievement in comparison with previous works where LDLC lattices~\cite{LDLC} of sizes $n=10000$ and turbo lattices~\cite{sakzadmanuscript} of size $n=10131$, at symbol error rate (SER) of $10^{-5}$, achieves $\alpha^2=0.8$ dB and $\alpha^2=0.5$ dB away from capacity respectively.


\section{Concluding remarks}~\label{ConclusionandSuggestionforLaterResearch}
The $1$-level regular LDPC lattices have been analyzed. It has been shown that these lattices are equivalent to the lattices constructed based on Construction A. It has been also established that these lattices are the same as LDLC lattices. Experimental results show that they outperform well-known LDLC lattices. That is due to the excellent performance of employed underlying LDPC codes. In summary, the $1$-level regular LDPC lattices perform very well among the regular LDPC lattices.

Further researches have to be done in order to find proper shaping methods. In that case one can extract good lattice codes from LDPC lattices which are appropriate for both Rayleigh fading and AWGN channels~\cite{viterbo}.

\end{document}